\documentclass[amsmath,amssymb,onecolumn, a4paper,14pt]{revtex4}
\usepackage{amsmath}
\usepackage{amssymb}
\usepackage{amsmath,amsthm,amssymb,amscd}
\usepackage{latexsym}
\usepackage{indentfirst}
\usepackage{subfigure}
\usepackage{graphicx}
\usepackage{extarrows}
\usepackage{bm}
\usepackage{multirow}
\usepackage{rotating}
\usepackage[colorlinks,linkcolor=blue,hyperindex,dvipdfm]{hyperref}

\date{\today}

\begin{document}

\title{Correlations between promoter activity and its nucleotide positions in  spacing region}

\author{Jingwei Li, Yunxin Zhang} \email[Email: ]{xyz@fudan.edu.cn}
\affiliation{Shanghai Key Laboratory for Contemporary Applied Mathematics, Centre for Computational Systems Biology, School of Mathematical Sciences, Fudan University, Shanghai 200433, China. }

\begin{abstract}
Transcription is one of the essential processes for cells to read genetic information encoded in genes, which is initiated by the binding of RNA polymerase to related promoter. Experiments have found that the nucleotide sequence of promoter has great influence on gene expression strength, or promoter activity. In synthetic biology, one interesting question is how we can synthesize a promoter with given activity, and which positions of promoter sequence are important for determining its activity. In this study, based on recent experimental data, correlations between promoter activity and its sequence positions are analyzed by various methods. Our results show that, except nucleotides in the two highly conserved regions, $-35$ box and $-10$ box, influences of nucleotides in other positions are also not neglectable. For example, modifications of nucleotides around position $-19$ in spacing region may change promoter activity in a large scale. The results of this study might be helpful to our understanding of biophysical mechanism of gene transcription, and may also be helpful to the design of synthetic cell factory.
\end{abstract}

%\keywords{}

\maketitle

\section{Introduction}
In cells, generic information is transcribed from DNA template to messenger RNAs (mRNAs) by RNA polymerase (RNAP) through a series of complex processes, called transcription. The key step that starts transcription is the binding of RNAP to a special nucleotide sequence in DNA, which usually lies to the upstream of transcription start site of the gene, and is called promoter \cite{Mulligan1984,Ross1993RNApromoter,Cheetham1999RNApromoter,Campbell2002}. Experimental data show that expression strength of corresponding gene, or protein production rate, is greatly influenced by the nucleotide sequence of promoter \cite{Hawley1983Nature,Ruhdal1998LibrarySpacer,Jensen1998}. Therefore, it is important for synthetic biology and genetic engineering to choose appropriate promoter to achieve needed expression strength. Meanwhile, experiments also find that the activity of a promoter does not change with genes it expresses \cite{Alper2005LibraryReuse,Mutalik2013Reuse,Cox2007Reuse,Barnard2007Reuse}. It means that a promoter with strong activity in expression of one gene can always  express other genes in relatively high strength.
Therefore, it is biologically meaningful to establish promoter libraries with promoter strength (or activity) changing in a large scale, as those have been experimentally done in \cite{Jensen1998,Mey2007LibraryNearby,Alper2005LibraryReuse,Rud2006Library,Rhodius2012,Lu2012,Wu2013}.
Meanwhile, in order to understand the regulation mechanism of promoter in gene expression, various nucleotide sequence dependent models are also designed, which  are expected to be applicable in real cells \cite{Kiryu2005Support,Rhodius2010OtherPredict,Kinney2010Regulatory,Brewster2012Plos,Li2014}.

For a constitutive promoter, i.e. its activity is not influenced by transcription factors, the initiation rate of transcription is mainly determined by RNAP binding rate to corresponding promoter. As almost all previous theoretical studies about promoter activity, this study assumes that the RNAP binding rate depends only on the nucleotide sequence of promoter. It has been found that, in {\it E. coli}, promoters include two highly conserved hexamers, which are usually called  $-10$ box and $-35$ box, and they are essential to promoter activity \cite{Hawley1983Nature,Ross1993RNApromoter,Ruhdal1998LibrarySpacer}. In some theoretical models, only nucleotides in these two regions, as well as the discriminator region and transcription start region, are considered in detail. While for the spacing region between $-10$ and $-35$ boxes, only the length of it is assumed to contribute to promoter activity as a model penalty term. %the spacing region between the $-10$ and $-35$ boxes is assumed to contribute to promoter activity only by its length as a model penalty term
\cite{Rhodius2010OtherPredict}. However, recent experimental data presented in \cite{Wu2013} show that promoter activity also changes with the nucleotide types in spacing region.

Although in some theoretical studies, nucleotides in the spacing region are also included explicitly, correlations between sequence positions in spacing region and  promoter activity are not discussed \cite{Jensen1998,Ruhdal1998LibrarySpacer,Kiryu2005Support,Kinney2010Regulatory,Brewster2012Plos}.
In this study, based on promoter libraries given in \cite{Wu2013,Mutalik2013Reuse}, these correlations are calculated by various methods. Here, large values of correlation mean that promoter activity changes greatly with nucleotide type at this position, while small values of correlation mean that promoter activity is insensitive to the nucleotide type at this position. Our results show that, except sequence positions around the $-10$ box and $-35$ box, nucleotides at positions $-20$, $-19$, and $-18$, which lie in the spacing region, play important roles in determining promoter activity. On the contrary, nucleotides at positions $-23$ and $-15$ seem to be of no significance.

\section{Results}\label{results}
The importance of nucleotide hexamers in $-35$ box and $-10$ box of a promoter has been discussed previously \cite{Dehaseth1998,Hawley1983Nature,Djordjevic2011}. The main aim of this study is to find which positions in spacing region are important for promoter activity. In other words, modifications of nucleotides in these positions may change expression strength greatly. To achieve this, a linear model is designed to describe the relationship between nucleotide sequence and promoter activity, which is based on basic principles of statistical physics and the assumption that the strength of gene expression is proportionate to RNAP binding rate to promoter (see Sec. \ref{methods}).

For the data from Wang's study group \cite{Wu2013}, promoter sequences are only different in spacing region, i.e. from position $-13$ to $-29$. So in our analysis, only nucleotides in spacing region are considered. The data obtained by Mutalik {\it et al} in \cite{Mutalik2013Reuse} consist of three groups, which are denoted by {\bf mpl}, {\bf rpl}, and {\bf pilot}. But in no matter which group of them, nucleotide hexamers in $-35$ box and $-10$ box are not fixed. Therefore, nucleotide sequence from position $-1$ to $-35$ are considered in our analysis. For each promoter, the length of its spacing region is 17.

The relationship between nucleotide sequence and promoter activity is described by three $k-$mer models with $k=1,2,3$. Here $k-$mer model assumes that promoter activity can be determined by all $k$ adjacent nucleotide groups, and at each sequence position $i$, there are altogether $5^k$ variables, which consist of all $k-$permutations of nucleotides A, T, G, C, and $-$, where \lq\lq$-$" means that the corresponding nucleotide is missed. Due to the large number of variables, partial least squares (PLS) regression is used in calculations. To improve model accuracy, 10-fold cross-validation is used in the PLS regression (see Sec. \ref{methods}).

For each group of experimental data of promoter strength obtained in \cite{Wu2013,Mutalik2013Reuse}, model coefficient values related to each sequence position can be obtained by PLS regression with 10-fold cross validation. We calculate their variance and range (defined as the difference between their maximum and minimum), which are regarded as two criteria of influence of the corresponding sequence position to promoter activity. The main difference between variance and range is that, variance is the average value of deviations of model coefficient values from their average
while the range only describes their variation range.

The main aim of this study is to analyze the correlation between sequence position, especially those in spacing region, and promoter activity by using all the four data groups in \cite{Wu2013,Mutalik2013Reuse}. But the scales of variances and ranges obtained from the four data groups with three versions of $k-$mer model are different. If operating on these values directly, the information in small scale values will be absorbed by that in large scale values, and therefore they influence will be weakened inappropriately. To avoid this, we turn all variances and ranges into so-called scores by following method. We sort positions from 1 to 35 (17 for  {\bf Wang}'s data) by their variances (or ranges) in descending order. The position ranked $n-$th is scored $36-n$ (scored $18-n$ for  {\bf Wang}' data). So far, for each data group, based on the variance and range of the three $k-$mer models, six scores can be obtained. For convenience, scores obtained by the variance and range of $k-$mer model are denoted by $V_k$ and $R_k$ respectively. At the same time, another score based on F-statistic can be obtained by the following idea. If promoter activity depends greatly on a position, then the precision of corresponding model with this position neglected will be low. So its 10-fold cross-validation error will be large. Using this 10-fold cross-validation error and by the same method as discussed above, one new score can be obtained for each sequence position, which is denoted by $F$ for simplicity. Note that, in calculations of score $F$, the 10-fold cross-validation errors are obtained by $1-$mer model.

For each group of data, we have altogether seven scores. For  {\bf Wang}'s data, the score of position in spacing region ranges from 1 to 17, while for other three groups of data, the score ranges from 1 to 35. To place all scores in the same level, linear transformations are applied to each of the 28 scores such that for each score, its minimal and maximal values corresponding to sequence positions in spacing region are 0 and 100 respectively. See Table SI-SIV in \cite{supplemental} for the processed data.

\begin{figure}
  \centering
  \includegraphics[width=15cm]{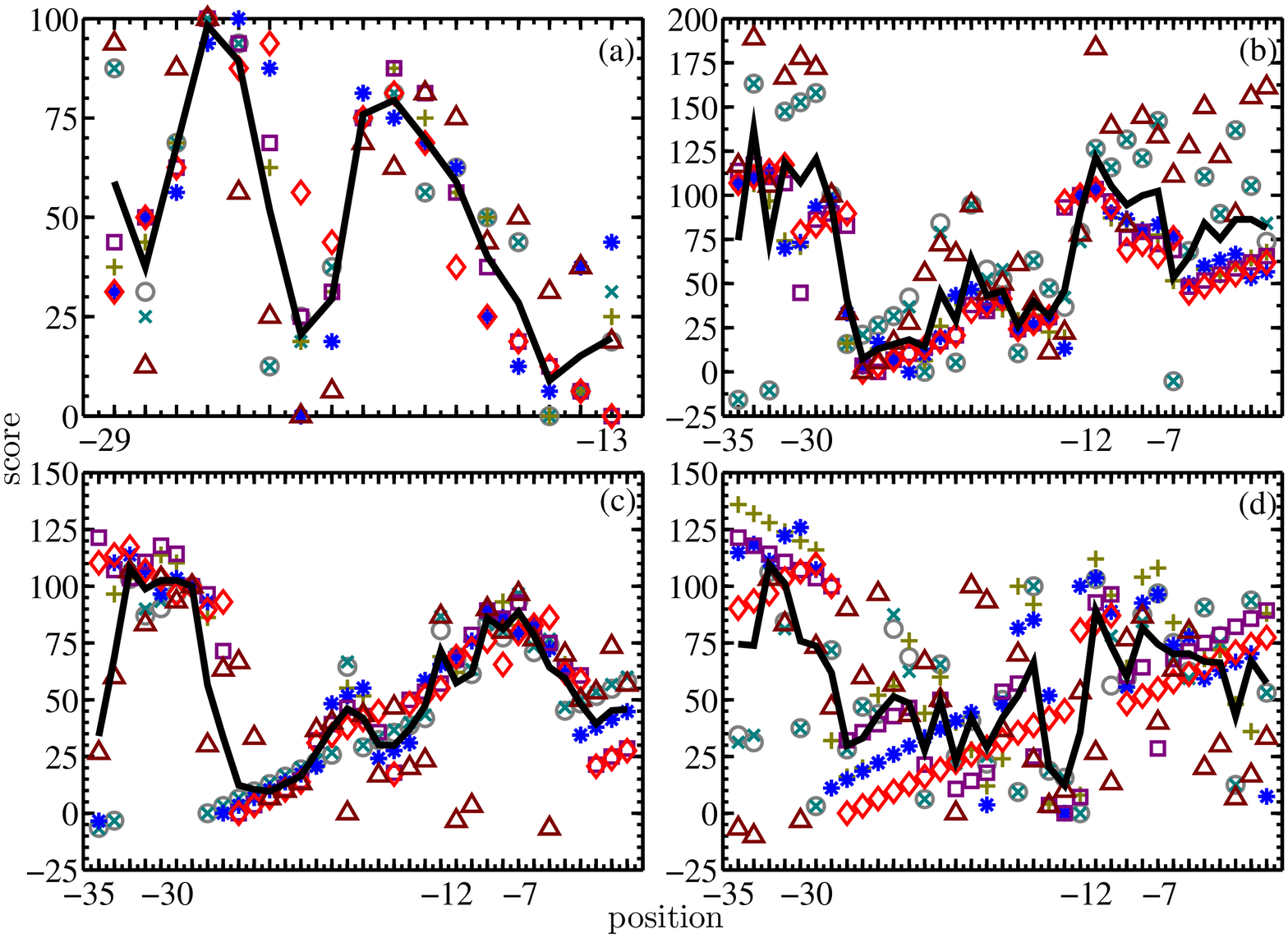}\\
  \includegraphics[width=13.8cm]{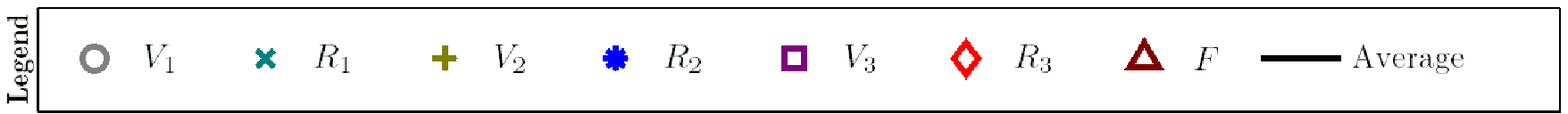}
  \caption{The normalized scores $V_k$, $R_k$ (for k=1,2,3), and $F$ for data group  {\bf Wang} presented in \cite{Wu2013}  {\bf (a)}, and data groups {\bf mpl}, {\bf rpl}, and {\bf pilot} presented in \cite{Mutalik2013Reuse} {\bf (b,c,d)}. The thick black line is the average of $V_k, R_k$ and $F$. The x-axis is nucleotide sequence position in promoter. Where nucleotide hexamer from position $-7$ to $-12$ is called $-10$ box, and nucleotide hexamer from position $-30$ to $-35$ is called $-35$ box. Scores $V_k$ and $R_k$ are obtained from the variance and the range of model coefficients in $k-$mer model, respectively. Score $F$ is obtained from F-statistic of model coefficients in $1-$mer model. For detailed score values, see Tables SI-SIV in \cite{supplemental}.
  }\label{MainFig1}
\end{figure}

Due to the differences of strains and measurement methods used in different experiments \cite{Wu2013,Mutalik2013Reuse}, correlations between sequence position and promoter activity obtained from different data may be different. The seven scores as well as their average obtained from each data group are plotted in Figs. \ref{MainFig1}{\bf (a-d)}, see also Table SI,SII,SIII,SIV for the score details \cite{supplemental}. Where x-axis shows the promoter sequence position, and y-axis shows the corresponding score. To distinguish the scores and their average, different markers and colors are used in Figs. \ref{MainFig1}{\bf (a-d)}, which are listed at the bottom of Fig. \ref{MainFig1}. The average scores obtained from data groups {\bf mpl}, {\bf rpl}, and {\bf pilot} measured in \cite{Mutalik2013Reuse}, see the thick black lines in Figs. \ref{MainFig1}{\bf (c-d)}, show that the $-35$ box and $-10$ box are the most important regions, while the spacing region (the region between $-35$ and $-10$ boxes)
is the least important region. Which is consistent with previous experimental observations.
The scores of spacing region positions near these two boxes are also high. For data groups  {\bf Wang}, {\bf mpl}, and {\bf rpl}, positions around $-19$ also have high scores. However, according to data {\bf Wang}, positions around $-26$ get higher scores than those around position $-19$. For data  {\bf pilot}, average score oscillates in spacing region and reaches its maximum at position $-17$. The average scores listed in Table SV (see \cite{supplemental}) show that, for data {\bf Wang}, the four most important positions in spacing region are $-26, -25, -21, -20$, and the four least important positions are $-15, -14, -13, -23$. For data  {\bf mpl}, the most important position in spacing region is $-20$, and the least important position is $-27$.
For data  {\bf rpl}, the most important positions in spacing region are $-19, -18$, while the least important positions are $-24, -25, -26, -23$. Finally, for data  {\bf pilot}, the most important positions in spacing region are $-16, -29$, and the least important positions are $-14$, $-15$ and $-21$.

\begin{figure}
  \centering
  \includegraphics[width=15cm]{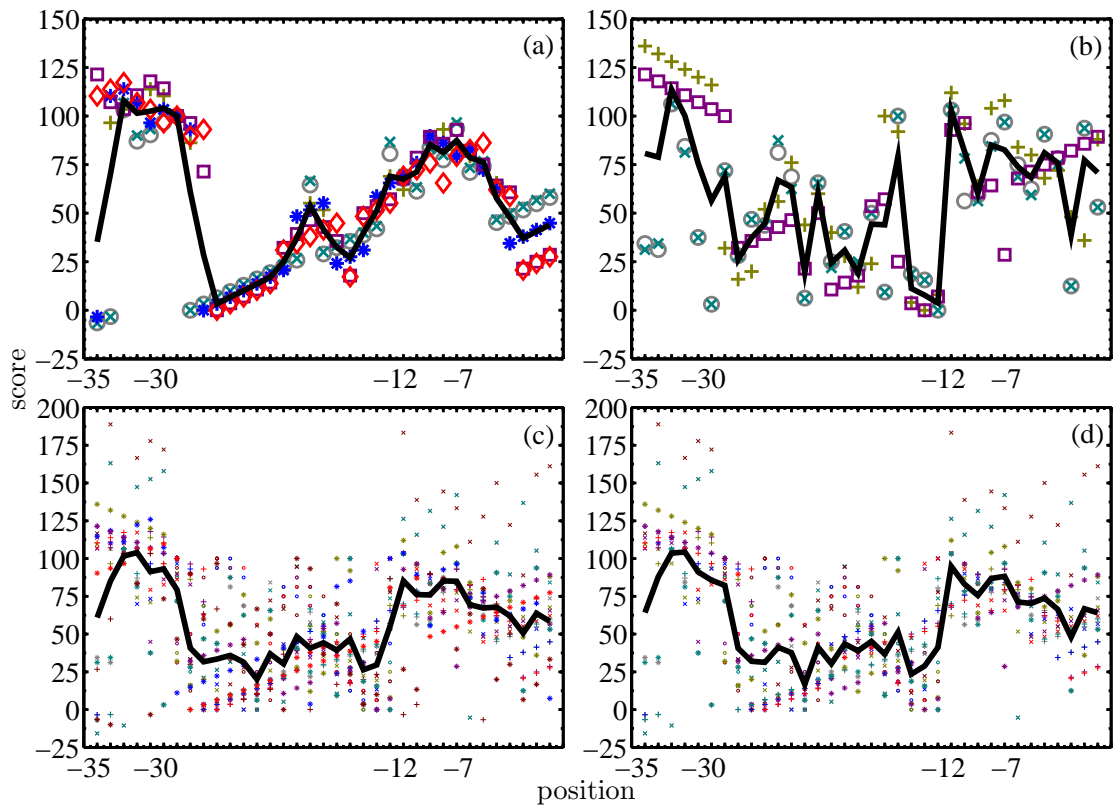}\\
  \caption{Survival scores after data clustering and their average for data groups {\bf rpl}  {\bf (a)} and {\bf pilot}  {\bf (b)} measured in \cite{Mutalik2013Reuse}, with the same legend as in Fig. \ref{MainFig1}. For data groups  {\bf Wang} and {\bf mpl}, all scores are survived after data clustering. See Fig. S1 in \cite{supplemental} for results of the data clustering.  {\bf (c)} All scores for the four data groups and their weighted average (thick black line), with sample numbers as the weights.  {\bf (d)} The survived 23 scores after data clustering and their weighted average. In  {\bf (c,d)}, scores for data groups  {\bf Wang}, {\bf mpl}, {\bf rpl}, and {\bf pilot} are marked by '$\circ$', '$\times$', '+', and '$*$' respectively. For detailed values of average scores, see Tables. SV and SVI in \cite{supplemental}. }\label{MainFig2}
\end{figure}

In the following, we will use three methods to put all the 28 scores together to find the most/least relevant positions in spacing region, which are expected to be generally true in any {\it E. coli} strain. One straight-forward method is to calculate the weighted average of the four average scores obtained from the four data groups, with sample numbers as the weights. The sample numbers for data  {\bf Wang}, {\bf mpl}, {\bf rpl}, and {\bf pilot} are 35, 69, 113, and 154 respectively. Since the sequence length of promoters in  {\bf Wang}'s data is different from those in data  {\bf mpl}, {\bf rpl}, and {\bf pilot}. We only average the four average scores in spacing region, i.e. from  position $-13$ to position $-29$. In other regions, the overall average is obtained by the three data groups  {\bf mpl}, {\bf rpl}, and {\bf pilot}. All the 28 scores and their average value are plotted in Fig. \ref{MainFig2}{\bf(c)}. These plots show that, generally, the $-10$ and $-35$ boxes are more important for determining promoter activity, while the spacing region is the least relevant region. From the average scores listed in Table SV (see \cite{supplemental}), we found that, except the two positions $-29, -13$ which are adjacent to $-35$ or $-10$ boxes, the most important positions in spacing region are $-20, -16$ and $-18$, while the least relevant positions are $-23$ and $-15$. For convenience, this method to calculate the average score is call weighted score method (WSM).

The second method to find the correlation of sequence position in spacing region to promoter activity is called four partitions method (FPM). Since for each data group, the sequence positions in spacing region are divided into four relevant groups, {\it important} group, {\it sub-important} group, {\it sub-unimportant} group, and {\it unimportant} group. In this method, sequence positions in spacing region are firstly sorted in descending order according to their average scores. Then the first five positions are assigned to the {\it important} group, and the rest 12 positions are assigned to the other three groups in turn, with four positions in each group, see Table SVII \cite{supplemental}. Then for each sequence position, we count the number of times it lies in the four relevant groups, see columns 2 to 5 in Table SIX \cite{supplemental}. According to this method, position $-29$, which is adjacent to the $-35$ box, is the most important position in spacing region because it is assigned to {\it  important} group three times and assigned to {\it  sub-important} group one time. The important positions in spacing region which are not adjacent to $-10$ or $-35$ boxes are $-20$ and $-19$, and the most unimportant positions are  $-23$ and $-15$.

The third method to discuss the correlation between sequence position and promoter activity is called signed-rank and rank-sum method (SRRSM). In which, according to the seven scores from each data group, the sequence positions are divided into three groups by Wilcoxon signed-rank test and Wilcoxon rank-sum test (see Sec. \ref{methods}), see Table SVIII in \cite{supplemental}. According to these tests, the score of any position in the first group is larger than that of any position in the third group with significance level $\alpha=0.05$. For any position in spacing region, the number of times that it lies in given group is listed in Table SX \cite{supplemental}. Which shows that the most important positions are $-29$, $-20$, $-19$, $-18$, and $-16$, while the most unimportant position is $-23$.

The reason that we used seven different ways to calculate position score for each data group is that, generally, sort results of sequence positions obtained by different scoring methods are not the same, but we cannot know which one of them is more reasonable. In previous discussion, we simply used their average to sort sequence positions in spacing region of promoter. Another method to deal with this problem is to use data clustering method to exclude the scores that are much different from others. Or in other words, we only keep and average the scores that give similar sort results, and exclude the ones that the sort results obtained by them are peculiar (see Sec. \ref{methods}). Since we are mainly interested in the relevant of sequence positions in spacing region, in data clustering process only scores of the spacing region positions are used. The results of clustering are plotted in Figs. S1 {\bf (a,b,c,d)} \cite{supplemental}. In this study, scores with clustering distance larger than 0.55 will be excluded. For data  {\bf Wang} and {\bf mpl}, no score is excluded. For data  {\bf rpl}, only the score $F$ is excluded. While for data  {\bf pilot}, only scores $V_1, V_2, V_3$, and $R_1$ are kept.

Through data clustering with distance criterion 0.55, there is no change for data  {\bf Wang} and {\bf mpl}. The new average scores of data  {\bf rpl} and {\bf pilot}, obtained by averaging only the survival scores (scores with distance less than 0.55), are plotted in Fig. \ref{MainFig2}{\bf (a)} and Fig. \ref{MainFig2}{\bf (b)} respectively. See also Table SVI in \cite{supplemental} for their detailed values. Fig. \ref{MainFig2}{\bf (a)} shows that, for data  {\bf rpl}, except the positions adjacent to $-10$ or $-35$ boxes, position $-19$ is the most important one, while position $-26$ is the most unimportant one. But for data  {\bf pilot}, the most important position is $-16$, and the most unimportant positions are $-13, -14, -15$. All the 23 survival scores and their weighted average, with sample numbers as weights, are plotted in Fig. \ref{MainFig2}{\bf (d)}. The sorted sequence positions in spacing region obtained by the data clustering method are listed in Table SVI \cite{supplemental}. This clustering method gives that, except the positions adjacent to $-10$ or $-35$ boxes, the most important and unimportant positions in spacing region are $-16, -18$, and $-23, -15$ respectively. This method is denoted by CWSM.

The four partitions method can also be modified by using data clustering process, which is denoted by CFPM. The new four partitions of the four data groups are listed in Table SVII (see \cite{supplemental}), and the number of times that sequence position lies in given partition is summarized in Table SIX \cite{supplemental}, which shows that positions $-29$, $-19$, and $-20$ are important while the positions $-23$ and $-15$ are unimportant. Similarly, the results of data clustering version of SRRSM (denoted by CSRRSM) are listed in Table SVIII and Table SX \cite{supplemental}. Where the results for data group {\bf pilot} are not presented since there are only four survival scores which are not enough to get reliable results from Wilcoxon signed-rank test or Wilcoxon rank-sum test. The results in Table SX show that positions $-29$, $-18$, $-19$, and $-20$ are more important, while position $-23$ seems to be unimportant \cite{supplemental}.

Finally, the importances of sequence positions in spacing region obtained from the six methods are summarized in Table \ref{Implist} \cite{supplemental}. It can be found that, in the region between position $-27$ and $-15$, positions $-20$, $-19$, and $-18$ play special roles in promoter activity. On the contrary, positions $-23$ and $-15$ seem to be of no significance. Here we neglect positions $-29$, $-28$, $-14$, and $-13$ because they are too closed to the $-10$ or $-35$ boxes.

\section{Methods}\label{methods}
\subsection{The $k-$mer models}
Let $y$ be gene expression strength. According to basic principles of statistical physics and the assumption that strength of gene expression is proportional to attachment rate of RNA polymerase (RNAP) to the upstream promoter, we have
\begin{equation}\label{model}
y=K\exp[-\Delta G/(k_BT)],
\end{equation}
where $\Delta G$ is the energy barrier of RNAP attachment to promoter. $k_B$ is the Boltzmann constant. $T$ is absolute temperature, and in this study, $T=300K$ (37$^{\circ}$C) is used. The constant $K$ depends on all other experimental conditions such as the concentration of RNAP, the speed of transcription elongation and termination, as well as the speed of the following translation process. Therefore, the value of $K$ will be different for data measured in different experiments.

We assume that the energy barrier $\Delta G$ can be completely determined by the nucleotide sequence of promoter \cite{Li2014}. The simplest way to establish this relationship is to assume that each sequence position contributes to $\Delta G$ independently and additively, and $\Delta G$ can be given by the following linear combination,
\begin{equation}\label{version1}
\Delta G=\sum_{i\in D}\Delta G_{i,b_i},
\end{equation}
in which $D$ is the set of sequence positions, and $b_i\in\{{\rm -,A,T,G,C}\}$ is the nucleotide at position $i$ ($b_i={\rm -}$ means the nucleotide at position $i$ is missing). For data group {\bf Wang} \cite{Wu2013},
\begin{equation}
D=\{i|i\in \mathbb{Z},-29\le i\le -13\},
\end{equation}
while for data groups {\bf mpl}, {\bf rpl}, and {\bf pilot} \cite{Mutalik2013Reuse},
\begin{equation}
D=\{i|i\in \mathbb{Z},-35\le i\le -1\}.
\end{equation}
Eqs. (\ref{model},\ref{version1}) are the so called $1-$mer model.

If the expression of $\Delta G$ is replaced by
\begin{equation}\label{version2}
\Delta G=\sum_{i,i+1\in D}\Delta G_{i,b_i,b_{i+1}},
\end{equation}
i.e. the total energy barrier is completely determined by all adjacent nucleotide groups of length 2, then we get the $2-$mer model. Finally, if
\begin{equation}\label{version3}
\Delta G=\sum_{i,i+1,i+2\in D}\Delta G_{i,b_i,b_{i+1},b_{i+2}}
\end{equation}
then the corresponding model is called $3-$mer model.

The $1-$mer model (\ref{model},\ref{version1}) can be reformulated as follows
\begin{equation}\label{model1}
\log y=\log K+\sum_{i\in D}\sum_{b\in\{{\rm -,A,T,G,C}\}}\delta_{i,b}[-\Delta G_{i,b}/(k_BT)],
\end{equation}
where $\delta_{i,b}$, for $b\in\{{\rm -,A,T,G,C}\}$, is defined as follows
\begin{equation}
\delta_{i,b}=\left\{
               \begin{array}{cc}
                 1 & {\rm if}\ b=b_i \\
                 0 & {\rm else} \\
               \end{array}
             \right..
\end{equation}
For each promoter sample, its nucleotide sequence corresponds to a vector   $(\cdots,\delta_{i,\textrm{-}},\delta_{i,\textrm{A}},\delta_{i,\textrm{T}},
\delta_{i,\textrm{G}},\delta_{i,\textrm{C}},\cdots)$. Values of $\log K$ and $(-\Delta G_{i,b}/(k_BT))$ can be determined from measured data through partial least square regression (PLSR). Similarly, from the $2-$mer model and $3-$mer model, we obtained
\begin{equation}\label{model2}
\log y=\log K+\sum_{i,i+1\in D}\sum_{b\in\{{\rm -,A,T,G,C}\}}\sum_{b'\in\{{\rm -,A,T,G,C}\}}\delta_{i,b,b'}[-\Delta G_{i,b,b'}/(k_BT)],
\end{equation}
and
\begin{equation}\label{model3}
\log y=\log K+\sum_{i\in D}\sum_{b\in\{{\rm -,A,T,G,C}\}}\sum_{b'\in\{{\rm -,A,T,G,C}\}}\sum_{b''\in\{{\rm -,A,T,G,C}\}}\delta_{i,b,b',b''}[-\Delta G_{i,b,b',b''}/(k_BT)],
\end{equation}
respectively. Where
\begin{equation}\label{model4}
\delta_{i,b,b'}=\left\{
               \begin{array}{cl}
                 1 & {\rm if}\ b=b_i\,{\rm and }\, b'=b_{i+1} \\
                 0 & {\rm else} \\
               \end{array}
             \right.,\
\delta_{i,b,b',b''}=\left\{
               \begin{array}{cl}
                 1 & {\rm if}\ b=b_i,\, b'=b_{i+1},\,{\rm and }\, b''=b_{i+2} \\
                 0 & {\rm else} \\
               \end{array}
             \right..
\end{equation}

Note that, the sample numbers of data groups {\bf Wang}, {\bf mpl}, {\bf rpl}, and {\bf pilot} are 35, 69, 113, and 154 respectively. But the number of unknown variables in the above $k-$mer models may be very larger. For example, the $1-$mer model for data group {\bf Wang} includes $5\times17=85$ variables, and the $3-$mer model for data groups {\bf mpl}, {\bf rpl}, and {\bf pilot} includes $5^3\times(35-2)=4125$ variables. This is why this study uses PLSR but not LSR as usual.

To avoid overfitting, the number of principal components in PLSR is determined by 10-fold cross-validation. The promoter samples in each data group are randomly divided into 10 groups for 100 times. For each given number of principal components, we calculated the cross-validation error of each division, and then take their average as the cross-validation error of this component number. The error we used in this study is given by $\|\log\hat{y}-\log y\|_2^2=\|\log\hat{y}/y\|_2^2$, where $\hat{y}$ is the predicted strength of gene expression. The number of principal components will be accepted when the corresponding cross-validation error reaches its minimal value. Finally, model coefficients are determined by all promoter samples in corresponding data group through PLSR with the previously determined principal component number.

\subsection{Scores corresponding to variance and range of model coefficients}
All scores of each sequence position are calculated from its related model coefficients. For the $1-$mer models, the related coefficients of position $i$ are the coefficients of $\delta_{i,b}$, for $b\in\{{\rm -,A,T,G,C}\}$. For the $2-$mer models, the related model coefficients for position $-35<i<-1$ (or $-29<i<-13$ for data {\bf Wang}) are the coefficients of $\delta_{i-1,b,b'}$ and $\delta_{i,b,b'}$, for $b,b'\in\{{\rm -,A,T,G,C}\}$. The related coefficients for positions $i=-35$ and $i=-1$ are the coefficients of $\delta_{-35,b,b'}$ and $\delta_{-2,b,b'}$, respectively. Similarly, for data group {\bf Wang}, the related model coefficients for positions $i=-29$ and $i=-13$ are the coefficients of $\delta_{-29,b,b'}$ and $\delta_{-14,b,b'}$, respectively. Finally, for the $3-$mer models, the related model coefficients for position $-34<i<-2$ (or $-28<i<-14$ for data {\bf Wang}) are the coefficients of $\delta_{i-2,b,b',b''}$, $\delta_{i-1,b,b',b''}$, and $\delta_{i,b,b',b''}$. The model coefficients related to position $i=-35$ are the coefficients of $\delta_{-35,b,b',b''}$. The ones related to position $i=-34$ are the coefficients of $\delta_{-35,b,b',b''}$ and $\delta_{-34,b,b',b''}$. The ones related to position $i=-2$ are the coefficients of $\delta_{-4,b,b',b''}$ and $\delta_{-3,b,b',b''}$, and the ones related to position $i=-1$ are the coefficients of $\delta_{-3,b,b',b''}$. For data group {\bf Wang}, the model coefficients related to positions $i=-29,-28, -14, -13$ can be obtained similarly. Here, $b,b',b''\in\{{\rm -,A,T,G,C}\}$. From these related model coefficients, the variance $V_k$ and range $R_k$ in $k-$mer model can be obtained as described in section \ref{results}.

\subsection{Scores corresponding to F-statistic}
For the $1-$mer model given in Eq. (\ref{model1}), if the contribution to promoter activity from the nucleotide at position $k\in D$ is excluded, it will become
\begin{equation}\label{model5}
\log y=\log K+\sum_{i\in D,i\neq k}\sum_{b\in\{{\rm -,A,T,G,C}\}}\delta_{i,b}[-\Delta G_{i,b}/(k_BT)].
\end{equation}
From this modified model, a new principal component number of optimal PLSR can be found, together with a new optimal cross-validation error. By ranking sequence positions according to the descending order of these optimal cross-validation errors, the score $F$ can then be obtained as described in Sec. \ref{results}.

\subsection{Wilcoxon signed-rank test and Wilcoxon rank-sum test}
Suppose that, for each data group, the seven scores of each sequence position $i$ are independent and identically distributed random variables, then we can use Wilcoxon signed-rank test and Wilcoxon rank-sum test to show if the scores of a given position are significantly larger than those of others under a given significance $\alpha$. In this study, $\alpha=0.05$ is used, and each pair of sequence positions is tested by these two methods. The difference between two positions is regarded to be significant if at least one of the two tests is significant. All sequence positions are then divided into three partitions such that any position in the first partition has significantly larger score than that of any one in the third partition. See Table SVIII in \cite{supplemental} for the results of partition. Finally, the importance of position can then be analyzed by the three partitions for the four data groups as described in Sec. \ref{results}.

\subsection{Data clustering}
For the seven score vectors of each data group (see columns 2-8 in Tables SI-SIV \cite{supplemental}), data clustering is performed by following methods. The distance between any two score vectors $x$ and $y$ is defined as $1-{\rm corr}(x,y)$, with ${\rm corr}(x,y)$ to be the correlation coefficient. Firstly, distance between any two score vectors is calculated. The two closest score vectors are then clustered and replaced by their average, which is considered as a new point but with weight 2 in the following calculation of average. Repeat this process until all score vectors are clustered together. In this study, two score vectors will be regarded to be in the same class if the distance between them is shorter than 0.55, see Fig. S1 in \cite{supplemental}. In other words, the correlation between them is larger than 0.45. Only score classes which include more than one score are consider, and the ones include only one score are excluded. Our calculations show that, for any one of the four data groups which used in this study, there exists only one effective data class. For such a special case, the above data clustering process is equivalent to an excluding process.

\section{Remark and discussion}
It seems that the largest and smallest (negative) coefficients in various $k-$mer models are also reasonable evaluation criterions of sequence position importance. But it is not the case. For example, when model coefficients related to one sequence position are all very large, the corresponding correlation between this position and promoter activity may be not significant if their variance is very small. Since for such cases, the promoter activity is not sensitive to the nucleotide type in this position. In fact, one of the main aims of this study is to find that, in the synthesis process of promotor, the nucleotide in which positions should be chosen more carefully to achieve needed promoter activity, and the influence of nucleotide type in which positions can be neglectable. In other words, a sequence position has strong correlation with promoter activity means that, if the nucleotide in this position is not chosen properly, the promoter activity may vary in a large scale.

Except the correlation between single sequence position and promoter activity, we have also tried to analyze the importance of adjacent sequence position groups, i.e. tried to find which position groups (with two or three adjacent positions) are important for determining promoter activity, and which ones are not. However, for these complex cases, only the $k-$mer models for $k=2,3$ can be used. Therefore, there are only four or even two kinds of scores for each data group, which are not enough to get reliable results about the position importance. Therefore, the corresponding results are not shown in the study.

To analyze the importance of sequence positions in spacing region, in the initial stage of this study, we have tried to group the promoters in data groups {\bf mpl}, {\bf rpl}, and {\bf pilot} by their $-35$ box and $-10$ box. Since after this pretreatment, in each group, the nucleotide sequences in $-35$ box and $-10$ box are the same, so the sequence positions in spacing region can be scored easily. However, after this grouping, sample numbers of each subgroups are usually too small (usually less than 10) to get reliable results. Meanwhile, it is also unreliable if we only consider the spacing region positions but neglect the difference in $-35$ and $-10$ boxes. Therefore, in this study, all sequence positions, including the ones in spacing region, $-35$ and $-10$ boxes, and the discriminator region, are scored simultaneously, though only the normalized ones in spacing region are finally used.

In summary, the correlation between promoter sequence positions in spacing region, especially from position $-27$ to $-15$, and its activity is discussed based on three $k-$mer models. From the data presented in \cite{Wu2013}, we found that position $-26$ in promoter sequence is the most important one to determine promoter activity. While the data groups {\bf mpl} and {\bf rpl} presented in \cite{Mutalik2013Reuse} show that position $-19$ is the most important one, and data group {\bf pilot} in \cite{Mutalik2013Reuse} shows position $-17$ is the most important. These differences may be caused by the different {\it E. coli} strains used in experiments. To find the most important/unimportant positions that might be generally true for any {\it E. coli} strain, three methods, WSM, FPM, and SRRSM, are used to integrate all the 28 scores obtained from four data groups. The results suggest that positions around $-19$ display strong correlations with  promoter activity, while the nucleotide type at position $-23$ is almost irrelevant to promoter activity. Meanwhile, three modified methods are also used in the analysis, in which scores are firstly clustered to exclude the peculiar ones. But similar results are obtained, see Table \ref{Implist} in \cite{supplemental}.

\begin{acknowledgments}
This study was supported by the Natural Science Foundation of China (Grant No. 11271083), and the National Basic Research Program of China (National \lq\lq973" program, project No. 2011CBA00804).
\end{acknowledgments}

\clearpage

%\bibliographystyle{unsrt}
%\bibliography{reference}

\clearpage

\begin{table}
\center
\begin{tabular}{|c|c|c|c|c|c|c|}
  \hline
              & WSM & CWSM & FPM & CFPM & SRRSM & CSRRSM \\\hline
  Important   & -20 -19 -18 -17 -16 & -16 & -20 -19 & -20 -19 & -20 -19 -18 -16 & -20 -19 -18 \\\hline
  Unimportant & -23 -15 & -23 -15 & -23 -15 & -23 -15 & -23 & -23 \\
  \hline
\end{tabular}
\caption{Summary of results obtained from six methods to find correlations between sequence positions of promoter and its activity. The positions which have strong correlation with promoter activity are listed in the first row, and those have weak correlations are listed in the second row. The positions adjacent to $-10$ or $-35$ boxes usually have strong correlations with promoter activity. Therefore, in this table only the positions between $-27$ to $-15$ are presented. }\label{Implist}
\end{table}

\clearpage

\end{document}